\documentclass{article}
\usepackage{amssymb}
\usepackage{caption}
\usepackage{amsfonts}
\usepackage{amsmath}
\usepackage{fancyhdr}
\usepackage{lastpage}
\usepackage{bbm}
\usepackage{booktabs}
\usepackage{xurl}
\usepackage{graphicx}
\usepackage{hyperref}
\usepackage[paperheight=11in,paperwidth=8.5in,portrait,left=2.5cm,right=2.5cm,top=2.5cm,bottom=2cm]{geometry}
\usepackage{empheq}
\usepackage{tabularx}
\usepackage[title]{appendix}
\DeclareCaptionLabelFormat{AppendixTables}{A.#2}
\DeclareCaptionLabelFormat{AppendixFigures}{A.#2}
\usepackage{pythonhighlight}
\usepackage{rotating}
\usepackage{multirow}
\usepackage{subcaption}
\usepackage{makecell}
\usepackage{subcaption}

\usepackage{algorithm}
\usepackage[noend]{algpseudocode}
\makeatletter
\def\BState{\State\hskip-\ALG@thistlm}
\makeatother

\title{Applications of Traveling Salesman Problem on the Optimal Sightseeing Orders of Macao World Heritage Sites with Real Time or Distance Values Between Every Pair of Sites}
\author{Kin Neng Tong, Iat In Fong, In Iat Li, Chi Him Anthony Cheng, 
	\and
	Soi Chak Choi, Hau Xiang Ye, Wei Shan Lee\footnote{email: \href{mailto:WSLEEemails}{weishan\_lee@yahoo.com}}
}
\date{Pui Ching Middle School\\ Macao Special Administrative Region, People's Republic of China.}
\begin{document}
	\maketitle
	\abstract{The optimal route of sightseeing orders for visiting every Macao World Heritage Site at exactly once was calculated with Simulated Annealing and Metropolis Algorithm(SAMA) after considering real required time or traveling distance between pairs of sites by either driving a car, taking a bus, or on foot. We found out that, with the optimal tour path, it took roughly 78 minutes for driving a car, 115 minutes on foot, while 117 minutes for taking a bus. On the other hand, the optimal total distance for driving a car would be 13.918 km while for pedestrians to walk, 7.844 km. These results probably mean that there is large space for the improvement on public transportation in this city. Comparison of computation time demanded between the brute-force enumeration of all possible paths and SAMA was also presented, together with animation of the processes for the algorithm to find out the optimal route. It is expected that computation time is astronomically increasing for the brute-force enumeration with more number of sites, while it only takes SAMA much less order of magnitude in time to calculate the optimal solution for larger number of sites. Several optimal options of routes were also provided in each transportation method. However, it is possible that in some types of transportation there could be only one optimal route having no circular or mirrored duplicates.}
	\cfoot{\thepage}
	\section{Introduction}
	Macao,$^{\cite{macao wiki}}$ one of the two special administrative regions of People's Republic of China, is a very famous historical city. In 2005, the UNESCO World Heritage Committee$^{\cite{macao historical center}}$ announced that the Historic Center of Macao was inscribed on the UNESCO World Heritage List. The Historic Center of Macao$^{\cite{wh site mo}}$ represents the architectural heritage of the city’s historical remains, including city squares, streetscapes, churches and temples, such as the ruins of St Paul's Church, Senado Square, A-Ma Temple and the Leal Senado Building, and many others. Statistics records$^{\cite{statista}}$ showed that there are roughly 30 million tourists a year visiting Macao during last decade. Even if some visitors are prone to casinos, there is significant portion of visitors who are more interested in the heritage sites. Therefore, looking for an efficient route for tourists to visit all the heritage sites remains an important issue for tourism management.\\ 
	Generally speaking, a global optimization problem$^{\cite{global optimization}}$ is difficult to solve. Some specific problems have already had promising regular ways to solve. For example, the knapsack problem, assignment problem, as well as the set-cover problem, may be solved by linear programming.$^{\cite{GKT}}$ Meanwhile, convex functions and some specific concave functions may be solved by nonlinear programming.$^{\cite{AMP}}$ Without doubt, the generic method on the optimization is still hard to find.\\
	After H Whitney$^{\cite{lawer}}$ proposed the traveling-salesman problem (TSP) in a seminar talk in Princeton in 1930s, there have been several attempts that have been trying to tackle the problem. For example, Flood$^{\cite{Flood}}$ provided the obvious brute-force algorithm to deal with the problem, while proposing that the aim of the algorithm was to find, for finitely many points whose pairwise distances are known, the shortest route connecting the points. In 1954, Dantzig et al.$^{\cite{Dantzig et al}}$ calculated a path with the shortest road distance for 49 cities out of the 48 states and Washingtond D.C. Bently$^{\cite{Bently}}$ proposed a fast algorithm for this problem in a geometric aspect. Moreover, some techniques in Machine Learning also were applied to TSP. For instance, Tarkov$^{\cite{Tarkov}}$ solved TSP with Hopfield Recurrent Neural Network(RNN). In addition, the Generic Algoritm with Reinforcement Learning$^{\cite{Liu and Zheng}-\cite{Ottoni}}$ was also applied to solving TSP. Also, for several countries, one may make use of smopy and networkx libraries in Python to create a GPS-like route plan, exploiting the Dijkstra's algorithm$^{\cite{Dijkstra's algorithm}}$ to find out the shortest path.\\   
	However, TSP study specifically for Macao World Heritage sites remains unknown. In our study, we first collected data from GOOGLE MAP about the least time or shortest distance between pairs of Macao World Heritage Sites by either driving a car, taking a bus, or on foot. Later, after comparing the efficiency between brute-force enumeration and SAMA, we used SAMA to search for the routes with optimal time or distance for tourists to go over all the world heritage sites without repeating any site. Finally, in certain transportation methods, we observed that there could be only one optimal route that is not circular or mirrored repetitive.
	\section{Theorems}
	Suppose the $N$ Macao world heritage sites are labeled as $[0,1,2,\dots,N-1]$. Also, let $E_{i,j}$ denote time or distance required when traveling between a pair of sites $i$ and $j$. For example, the straight distance between site $i$ and site $j$ is $E_{i,j}=|\vec{r_{i}}-\vec{r_{j}}|$, where $\vec{r_{i}}$ and $\vec{r_{j}}$ are the respective coordinates of the two sites. We describe the problem as looking for a specific tour orders of $N$ such that the total traveling time or distance
	\begin{center}
		\begin{equation}\label{summation}
			E\bigl(\{\vec{r_{i}};\hspace{3pt}\forall i\}\bigr)=\sum_{i}\sum_{j<i}E_{i,j}
		\end{equation}
	\end{center}
	is minimized. 
	We may enumerate all possible site orders, and for each route of sightseeing orders we calculate the total distance or time, among which we could find out the minimum. The other possibility could be that we may make use of the algorithm of Traveling Salesman Problem$^{\cite{Newman}}$, for which the simulated annealing and Metropolis algorithm were used.
	\subsection{Brute-force enumeration without circular or mirrored duplicates}\label{brute-force enumeration}
	Because of the theory of permutation, it is easy to calculate, for the case of $N$, the number of all possibilities $N\{\vec{r_{i}};\hspace{3pt}\forall i\}$ that have no either circular or mirrored repetitions is  
	\begin{center}
		\begin{equation}\label{number of possible routes} 
			N\{\vec{r_{i}};\hspace{3pt}\forall i\}=\frac{N!}{2\times N}.
		\end{equation}
	\end{center}
	One approach to generate all possibilities could be Sawada's algorithm$^{\cite{Sawada}}$. Nevertheless, because our case is k-ary, it would be a lot more straight-forward to consider the following algorithm. For a given number $N$ that establishes a list of $[0,1,2,\dots,N-1]$, the pseudo-codes in Algorithm $\ref{alg:python snippet}$$^{\cite{necklace}-\cite{without reverse duplicates}}$ shows how we generated a list that constructed all possibilities of orders that have no repetitive circular or mirrored orders.

\begin{algorithm}
	\begin{algorithmic}[1]
		\Procedure{FINDORDERS}{}
		\State \textit{List} $\gets [0,1,2,\dots,N-1]$
		\State \textit{orders} $\gets$ \textit{empty list}
		\State $i \gets$ \textit{permutations in} $[1,2,\dots,N-1]$
		\BState \emph{loop}:
		\If {$ i\le$ \textit{reverse of} $i$} 
		\State \textit{orders} $\gets$ \textit{orders} + $i$
		\EndIf
		\State \textbf{goto} \emph{loop}
		\State \While{$term \in orders$} 
		\State \textbf{generate} \textit{first term in List} + $term$ 
		\EndWhile
		\EndProcedure
	\end{algorithmic}
	\caption{Procedure generating the k-ary necklace $[0,1,2,\dots,N-1]$ without duplicates of circular or mirrored repetitions.}\label{alg:python snippet}
\end{algorithm}
	For instance, calling the above procedure with $N=5$, $\mathbf{FINDORDERS(5)}$, would generate a list of permutations with length $N\{\vec{r_{i}};\hspace{3pt}\forall i\}=\frac{5!}{2\times 5}=12$
	without circular or mirrored duplicates. It should be realized that it is only a sufficient condition for routes being circular or mirrored duplicates to have same optimal $E$ in Eq. $\ref{summation}$; it is possible for routes not being repetitive on circular or mirrored orders to have same optimal value of $E$. \\ 
	It would be unrealistic to implement the above method for very large value $N$. If $N=25$, there could be $\frac{25!}{2\times 25}\approx 3.102\times 10^{23}$ such different values of $E$ to calculate, demanding unbearable computation time and memory. Because of this, a way of conquering this issue, Simulated Annealing and Metropolis Algorithm, comes into play.
	\subsection{Simulated Annealing and Metropolis Algorithm(SAMA)}\label{theory of SAMA}
	The first idea of is Simulated Annealing$^{\cite{kirkpatrick}}$ could be based on the fact that \textit{minimizing a mathematical formula, such as} Eq. $\ref{summation}$, \textit{is comparable to looking for a minimum energy state of a system in nature}. There is a caution, however, in the nature of work with hot materials, known for a long time. In order to get a good crystallization, we need to cool down the material slowly, finding out its energy configuration of global minimum. On the other hand, if we cooled down the material too quickly, we might got glassy solids because we only reached the energy configuration of the “local” minimum for the material. Second, the nature of simulated annealing has the stochastic components, making it possible to assist in the asymptotic convergence analysis$^{\cite{Emile}}$.\\
	First, we randomly generate an initial configuration of our system, then calculate the initial value of the quantity $E_{i}$. Then, because the system is \textit{ergodic}, we randomly modify a little bit the configuration, calculating again the quantity after the change, say $E_{j}$. If the new quantity is smaller than the old one, we know that we find out a new configuration with smaller value of the quantity for which we would like to minimize. However, if the new quantity is larger than the old one, we may not be able to carelessly reject the new configuration. Because in this way we may “cool down” our system too quickly, possibly falling into the local minimum. The idea of Metropolis Algorithm is that when the new quantity in the new configuration is larger than the old one, instead of absolutely rejecting the new configuration, we make it possible to accept the new configuration by the following Metropolis probability:$^{\cite{Newman}}$
	\begin{center}
		\begin{equation}\label{Metropolis probability} 
			P_{\beta}(E_{j})=\left\{
			\begin{array}{ll}
				\hspace{20pt}1\hspace{35pt}\mathrm{if\hspace{5pt}}  E_{j}\leq E_{i}; \\
				e^{-\beta (E_{j}-E_{i})}\hspace{15pt}\mathrm{if\hspace{5pt}} E_{j}>E_{i}.\\
			\end{array}
			\right.
		\end{equation}
	\end{center}
	In practice, while randomly generating a number z between 0 and 1, we decide to accept the new configuration, with aid of Eq.($\ref{Metropolis probability}$), if the random number $z$ satisfies 
	\begin{center}
		\begin{equation}\label{random number z} 
			z\leq e^{-\beta (E_{j}-E{i})}; 
		\end{equation}
	\end{center}
	otherwise we reject the swap and return to the last tour order. To be more specific, throughout our study we have chosen the maximum temperature $\mathrm{T_{max}}$, minimum temperature $\mathrm{T_{min}}$, and $\mathrm{\tau=k_{b}T}$ as in Table $\ref{tab:parameters}$.
	\begin{center}
		\begin{table}[htbp]
			\centering
			\begin{tabular}{ccc}
				\hline\hline
				$\mathrm{T_{max}}$ & $\mathrm{T_{min}}$  & $\mathrm{\tau=k_{b}T}$ \\\hline
				$1.0$              & $1\times 10^{-2}$   & $1\times 10^{3}$       \\
				\hline\hline
			\end{tabular}
			\caption{Parameters used in SAMA.}
			\label{tab:parameters}
		\end{table}
	\end{center}
	Theoretically we may demonstrate that it is possible to find out the global minimum under infinite number of iterations.$^{\cite{Emile}}$ But it is not realistic to iterate infinitely number of times. Therefore, we need to compensate between the optimized value and time we need to consume. It is worth mentioning that in the optimization problem it is quite impossible to assure that we already reached the global minimum. All we can do is to find out another new solution to see if the new one out-reaches a smaller value compared with the old one.
	\subsection{Locations of Macao world heritage sites}\label{locations of coordinates}
\begin{center}
		\begin{figure}[htbp]
			\centering
			\includegraphics[width=0.8\textwidth]{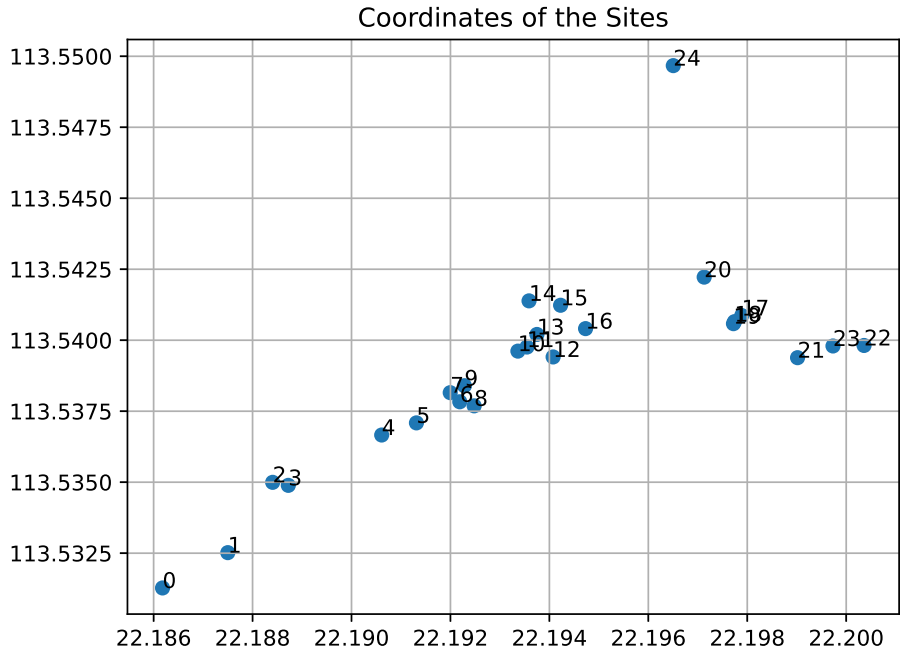}
			\caption{Coordinates of sites with the Site ID number used throughout the paper, indicating name of every site in Table A.$\ref{tab:site data}$. Site 17, 18, and 19 are too close to distinguish from one another. }
			\label{fig:coordinates}
		\end{figure}
	\end{center}

	We collected our data about coordinates on latitudes and longitudes of the Macao World Heritage Sites from GOOGLE MAP, tabulated in Table A.$\ref{tab:site data}$, with indications of names$^{\cite{macaoTO}}$. Figure $\ref{fig:coordinates}$ depicts the plot of site locations. Haversine formula$^{\cite{Haversine formula}}$ was used to convert from latitudes and longitudes to kilometers for a pair of sites.
	\section{Results and Discussions}
	We first compared the computation time between the brute-force method and SAMA. Later, for the first trial of SAMA, we introduced the latitudes and longitudes of the Macao World Heritage sites and calculated the optimal route with the minimum total distance by assuming that any pair of sites could be reached by a straight line. But this assumption is not realistic because in practice the real roads, streets or avenues in Macao are mostly not straight lines. Therefore, taking into consideration of real situations, for every pair of sites, we searched on GOOGLE MAP for the genuine time or distance required by either driving a car(Table A.$\ref{tab:car time}$ or Table A.$\ref{tab:car distance}$), taking a bus(Table A.$\ref{tab:bus time}$), or walking across the streets by pedestrians(Table A.$\ref{tab:pedestrian time}$ or Table A.$\ref{tab:pedestrian distance}$). Notice that GOOGLE MAP does not provide with data of distance for bus between pairs of sites. Data were tabulated in Appendix B to Appendix F. Through Table A.$\ref{tab:car time}$ to Table A.$\ref{tab:pedestrian distance}$, number of Site ID at the left-hand column refers to the departure site, whereas number of Site ID on the top row refers to the destination site.   
	\subsection{Comparison of computation time between the brute-force enumeration and SAMA on Fictitious Site Coordinates}\label{comparing comp time}
\begin{center}
		\begin{figure}[htbp]
			\centering
			\includegraphics[width=0.7\textwidth]{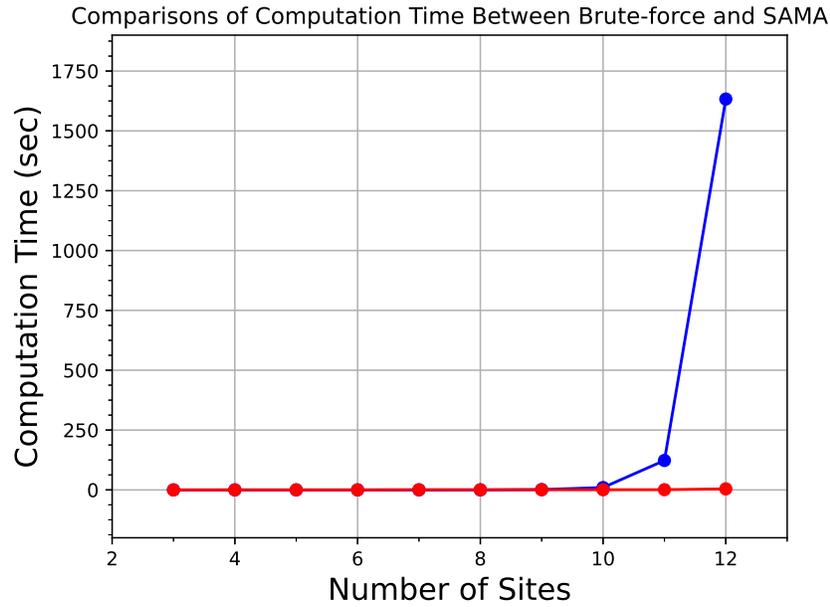}
			\caption{Comparison of Computation Time between Brute-force enumeration(blue curve) and SAMA(red curve). Computer specs: Intel Core i7-4500U CPU @ 1.80 GHz 2.39 GHz. RAM 8GB. 64-bit operating system and x64-based processor.}
			\label{fig:comparisons}
		\end{figure}
	\end{center}

	In order to compare time required by brute-force method and SAMA, we first randomly generate fictitious 12 sites, whose coordinates are lists in Table $\ref{tab:fictitious coordinates}$. Table $\ref{tab:comparisons brute force and SAMA}$ showed results of required computation time, after taking the average value on five times for each number of sites, under the condition of same shortest distance. It was evident that computation time for SAMA remained roughly less than 1 seconds for number of sites $N$ less than 11, while for $N=12$, 4.28 seconds. On the other hand however, for the brute-force method, for fewer number of sites, the required time was less than that of SAMA, referring to the fact that for fewer number of sites, the brute-force method prevails SAMA. But the advantage of SAMA gradually appeared for larger $N$ at least for two aspects. First, for $N=12$, the brute-force method required roughly $1600$ seconds to compute, while SAMA only required 4.28 seconds. The other advantage was that for larger number of $N$, brute-force method demands a lot of computer memory, causing it impractical to implement in reality. Meanwhile, it is not a problem for SAMA because of the characteristic nature in randomly selecting the state of the system to compute. Figure $\ref{fig:comparisons}$ shows plots of required time vs. number of sites. It is obvious that demanding time for brute-force method increased dramatically for larger number of sites.  Codes may be obtained via Ref.$\cite{github brute-force vs SAMA}$.

\begin{table}
	\parbox{0.4\linewidth}{
	\centering
	\begin{tabular}{@{}cll@{}}
		\hline\hline
		\multicolumn{1}{l}{Sites ID} & \multicolumn{1}{c}{X} & \multicolumn{1}{c}{Y} \\ \midrule
0 & 0.428919706 & 0.361347607 \\
1 & 0.530081873 & 0.698859005 \\
2 & 0.635617396 & 0.45383111 \\
3 & 0.164133133 & 0.584142242 \\
4 & 0.255455144 & 0.797792439 \\
5 & 0.700364502 & 0.329273766 \\
6 & 0.38266615 & 0.280563838 \\
7 & 0.524974098 & 0.349548083 \\
8 & 0.24761384 & 0.628859237 \\
9 & 0.30558778 & 0.083368041 \\
10 & 0.816248668 & 0.079397349 \\
11 & 0.413529397 & 0.340468561 \\
		\hline\hline
	\end{tabular}
	\caption{Fictitious coordinates of sites.}
	\label{tab:fictitious coordinates}
	}
	\hfill
	\parbox{0.6\linewidth}{
		\centering
	\begin{tabular}{*{4}{c}}
		\hline\hline
		number of Sites & $\star$ &  $\circledcirc$ & $\circledast$ \\ \hline
		3 & 0.84557967 & $\sim 0$ & $\sim 0$ \\
		4 & 1.22278990 & $\sim 0$ & 0.25484419 \\
		5 & 1.36353451 & $\sim 0$ & 0.25962210 \\
		6 & 1.55080373 & 0.00199866 & 0.23278880 \\
		7 & 1.67189785 & 0.01299381 & 0.76898718 \\
		8 & 1.67189945 & 0.11329699 & 0.61911297 \\
		9 & 1.70336830 & 1.37156701 & 0.81496930 \\
		10 & 2.06140792 & 9.85908175 & 0.64583468 \\
		11 & 2.55755845 & 122.93096805 & 0.93948197 \\
		12 & 2.55779639 & 1632.98429966 & 4.27968502 \\
		\hline\hline
	\end{tabular}
	\caption{Comparisons of computation time, average on five times for each number of sites, between brute-force enumeration and SAMA. $\star$ shortest distance by either  brute-force enumeration, or SAMA (a.u.). $\circledcirc$ computation  time by brute-force enumeration (sec). $\circledast$ computation time by SAMA (sec). Computer Specs were given in Figure $\ref{fig:comparisons}$}
\label{tab:comparisons brute force and SAMA}
	}
\end{table}
	\subsection{SAMA for two sites connected by a straight line}
\begin{center}
		\begin{figure}[htbp]
			\centering
			\includegraphics[width=0.7\textwidth]{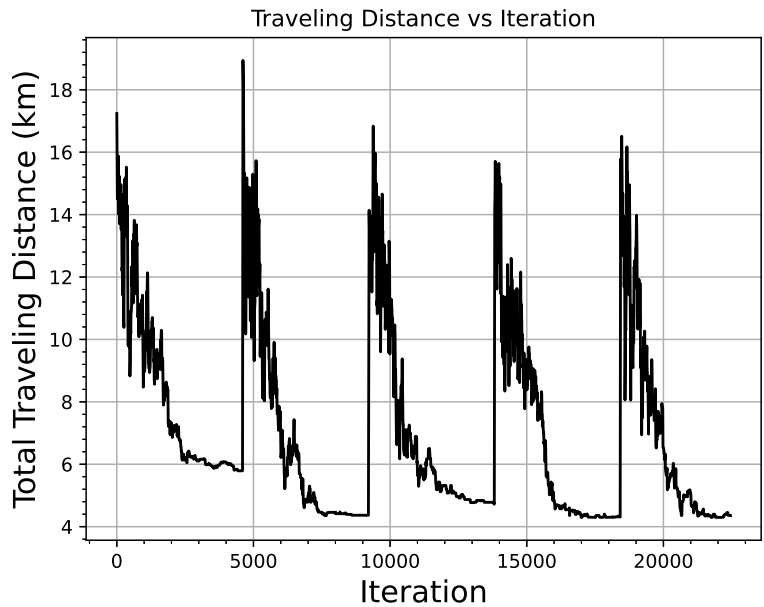}
			\caption{Total traveling distance (km) vs. Iteration. For each loop of SAMA, the system was cooled down to the relatively flat curve, and then we initiated another state of system to repeat SAMA. }
			\label{fig:distance vs iteration}
		\end{figure}
	\end{center}

\begin{center}
		\begin{figure}[htbp]
			\centering
			\includegraphics[width=1.0\textwidth]{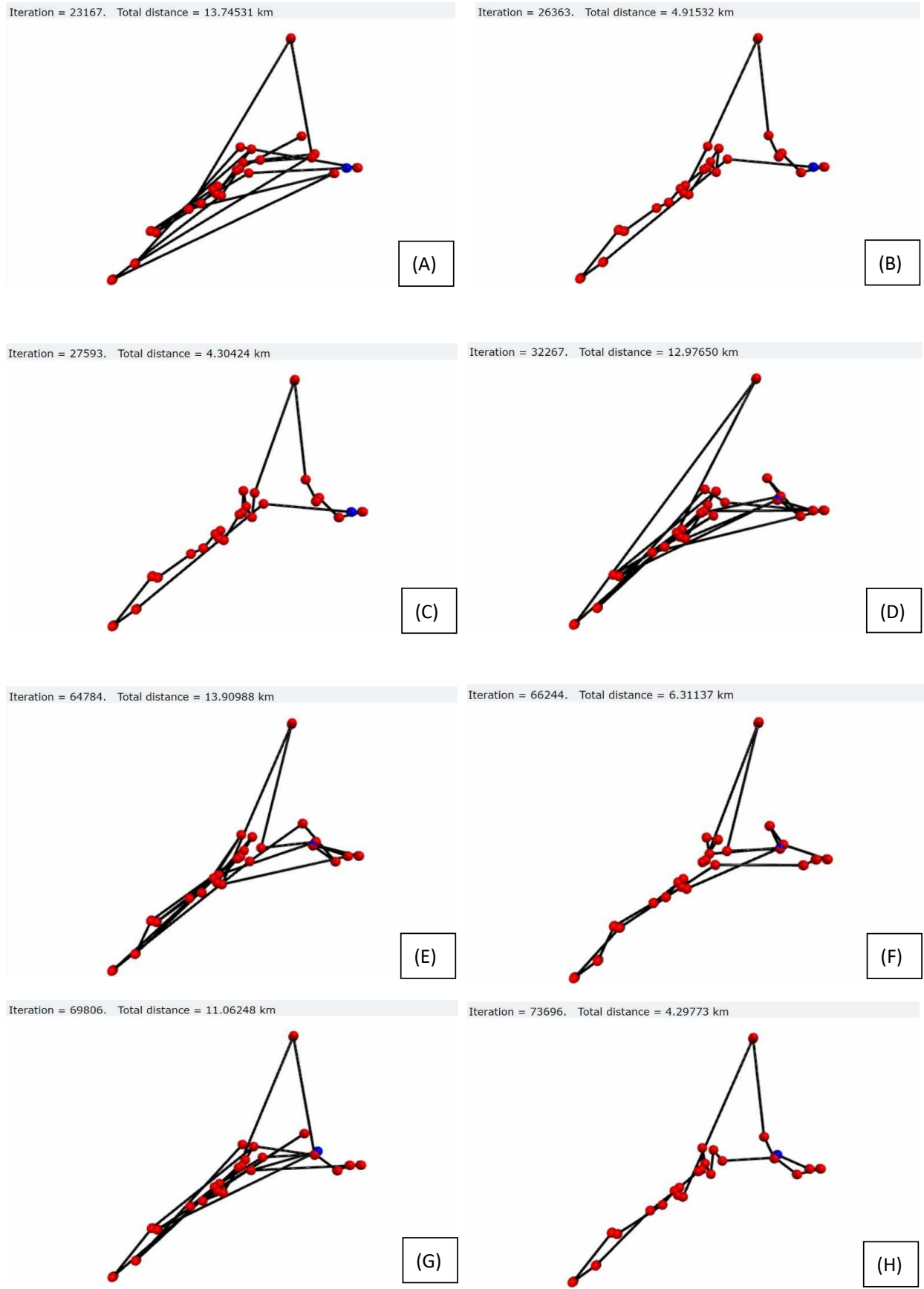}
			\caption{Snapshots of sightseeing order in the animations with various iterations, showing the total traveling distance at the specific iteration. The blue circle indicated the site of origin and end for the particular route.}
			\label{fig:animation}
		\end{figure}
	\end{center}

	First, we randomly chose a site to be the origin and the end, initiating a tour path with a (perhaps high) value of total traveling distance. Afterward, SAMA cooled down the system, trying to find out a route with shorter total distance. Figure $\ref{fig:distance vs iteration}$ showed the curve of total traveling distance vs. iteration. Keeping in mind that instead of absolutely rejecting all states with higher energy, SAMA also allows a state with higher energy described by the rule in Eq.$\ref{Metropolis probability}$, which is the reason for SAMA to put out the noisy curve within the iteration range $0$ to $3000$. Within Iteration $3000$ to $5000$, the curve remained relatively flat with the total distance roughly $6$ km. Nevertheless, this could only be a local minimum for the particular initial condition.\\
	In order to search for the global optimization, we started all over again by randomly initializing another starting point. This was shown at Iteration $4750$ with a dramatically increase of the total traveling distance. SAMA would cool down the system again to reach another (probably) local minimum. We repeated the process until a targeted value of total distance was achieved. Codes may be obtained via Ref.$\cite{SAMAV2_2}$.
	Figure $\ref{fig:animation}$ showed the snapshots of animation for some specific Iterations. The blue circle referred to the origin and the end for the specific route. In Figure $\ref{fig:animation}$(A), for Iteration equal to $23167$ with total distance $13.74$ km, the snapshot showed messy connections among pairs of sites, meaning that the tour order has not yet been optimized. Immediately after this, at Iteration equal to $26363$ with total distance $4.915$ km, $\ref{fig:animation}$(B) showed a much more organized route, which would later get improved in Figure $\ref{fig:animation}$(C) with a smaller value of total distance $4.3$ km.\\
	Later, as shown in Figure $\ref{fig:animation}$(D) to Figure $\ref{fig:animation}$(F), we chose another site as the starting point and the end point. From Iteration $32267$ to Iteration $66244$, the total distances were reduced from $12.977$ km to $6.311$ km. Furthermore, we chose another site as the blue circle, as in Figure $\ref{fig:animation}$(G). Finally we reached to the targeted total distance $4.29773$ km, acquiring the optimal route in Figure $\ref{fig:animation}$(H) at Iteration $73696$. Then the algorithm stopped. The whole animation video may be obtained in Ref$\cite{SAMAV2_2_animation}$.\\
	
	\subsection{Optimal time or distance for all possible types of transportation}
	In order to calculate the optimal route with genuine least time or shortest distance, we collected, for each transportation method, the true value of distance or time between a pair of sites with the form of matrices presented from Table A.$\ref{tab:car time}$ to Table A.$\ref{tab:pedestrian distance}$ in Appendix B to Appendix F. It is worth nothing that the matrices in above Tables may not be symmetric, demonstrating the fact that some types of transportation have different routes between the same pair of sites. Afterward, we tabulated, in Table A.$\ref{tab:ordersCarTime}$, Table A.$\ref{tab:ordersPedestrianTime}$, and Table A.$\ref{tab:ordersThreeTogether}$ in Appendix G, several routes that achieved optimal time or distance for every type of transportation. It should be understood without further notification that for each route, the circular or mirrored repetitions were also the optimal ones. In addition, in each type of transportation, the orders of optimal routes are either very similar or being circular or mirrored duplicates. For example, Table A.$\ref{tab:ordersBusTime}$, Table A.$\ref{tab:ordersPedestrianDistance}$, and Table A.$\ref{tab:ordersCarDistance}$ show that in every of these three kinds of transportation, there is only one order of route that is not circular or mirrored duplicates. On the contrary, in Table A.$\ref{tab:ordersCarTime}$ and Table A.$\ref{tab:ordersPedestrianTime}$, even if there are various optimal routes in these two types of transportation, the routes are quite similar in each kind of transportation.\\
	At last, in Table $\ref{tab:comparison all transportation}$ we may observe that it takes almost the same time for taking a bus (117 min), referring to Table A.$\ref{tab:ordersBusTime}$, and on foot (115 min), referring to Table A.$\ref{tab:ordersPedestrianTime}$. In spite of this, it takes shorter distance on foot (7.844 km), referring to Table A.$\ref{tab:ordersPedestrianDistance}$, than by driving a car (13.916 km), referring to Table A.$\ref{tab:ordersCarDistance}$. Codes may be retrieved via Ref$\cite{SAMAV3_1}$. These results suggest that the traffic condition in Macao imminently acquired to improve. The average computation time was taken by calculating the mean computation time of five routes in each transportation. Time demanded for computation ranged from 5 minutes (for least time on foot) to 74 minutes (for shortest distance by car). There seemed no clear pattern or tendency in types of transportation for the reason why required time for computing the optimal routes were different. 
	\begin{table}[htbp]
		\centering
\begin{tabular}{@{}ccccc@{}}
	\toprule\hline
	\multicolumn{1}{l}{\multirow{2}{*}{\makecell{Types of\\ Transportation}}} & \multicolumn{1}{l}{\multirow{2}{*}{Least time (min)}} & \multicolumn{1}{l}{\multirow{2}{*}{Shortest distance (km)}} & \multicolumn{2}{l}{\begin{tabular}[c]{@{}c@{}}Average computation time for\\the optimal value (min : sec)\end{tabular}} \\ \cmidrule(l){4-5} 
	\multicolumn{1}{l}{} & \multicolumn{1}{l}{} & \multicolumn{1}{l}{} & Least time & Shortest Distance \\ \midrule
	Car & 78 & 13.916 & 10:34 &  74:25\\
	Bus & 117 & Not Available & 44:39 & Not Available \\
	Pedestrian & 115 & 7.844 & 5:19 & 33:25\\ 
	\hline\hline
\end{tabular}
		\caption{Comparison of all types of transportation, including driving a car, taking a bus, or on foot. Computer Spec is given in the caption of Figure $\ref{fig:comparisons}$}
		\label{tab:comparison all transportation}
	\end{table}
	\section{Conclusions}
	We made use of Simulated Annealing and Metropolis Algorithm(SAMA) to obtain optimal sightseeing orders for least time or shortest distance of Macao World Heritage Sites with real time or distance between a pair of sites. Without repeating any site while completing a loop of the historical remains in Macao, the optimal least time for driving a car could be 78 min, while taking a bus would be 117 min, and 115 min on foot. On the other hand, the optimal shortest distance by car would be 13.916 km, while on foot, 7.844 km. This manifests the terrible traffic condition in the city of Macao, where improvement on public transportation is imperative.\\
	Also, we provided a simple algorithm to generate the k-ary necklace. Based on this we calculated computation time required in the brute-force method to obtain the optimal time or distance by enumerating all possible routes. Whereas for small number of sites, the brute-force enumeration performed much faster in calculating the optimal value, SAMA prevailed when number of sites increased. In our study, we demonstrated that when number of sites was 12, SAMA only required 3 orders of magnitude less in time than the brute-force method to obtain the optimal total distance.\\
	Last but not least, we provided several optimal routes in different kinds of transportation for tourists in Macao, from which choices may be made to manage their visiting plans more efficiently.  
	\section{Acknowledgment}
	We thank Pui Ching Middle School in Macao PRC for the kindness to support this research project. 
	
\newpage
\section{Appendix}
	\begin{appendices}
		\renewcommand\thetable{\arabic{table}} 
		\setcounter{table}{0} 
		\captionsetup{labelformat=AppendixTables}
		\renewcommand\thefigure{\arabic{figure}} 
		\setcounter{figure}{0} 
		\captionsetup{labelformat=AppendixFigures}
	\section{Locations of Macao World Heritage Sites}
	\begin{center}
		\begin{table}[htbp]
			\centering
			\begin{tabular}{@{}clll@{}}
				\hline\hline
				\multicolumn{1}{l}{Site ID} & \multicolumn{1}{c}{Latitude} & \multicolumn{1}{c}{Longitude} & \multicolumn{1}{c}{Name} \\ \midrule
				0 & 22.18618015 & 113.5312755 & Templo de A-Ma \\
				1 & 22.18749972 & 113.5325193 & Quartel dos Mouros \\
				2 & 22.1884061 & 113.5349956 & Largo do Lilau \\
				3 & 22.1887227 & 113.5348891 & Casa do Mandarim \\
				4 & 22.19061015 & 113.5366617 & Igreja de S\~tilde{a}  o Louren\c{c}o \\
				5 & 22.19131257 & 113.537089 & Igreja do Semin\'{a}rio de S\~{a}o Jos\'{e} \\
				6 & 22.19218864 & 113.5378323 & Largo de Santo Agostinho \\
				7 & 22.19199551 & 113.5381535 & Teatro Dom Pedro V \\
				8 & 22.19247942 & 113.5376891 & Biblioteca Sir Robert Ho Tung \\
				9 & 22.19228218 & 113.5383978 & Igreja de Santo Agostinho \\
				10 & 22.19336355 & 113.5396164 & Instituto para os Assuntos Municipais \\
				11 & 22.19354689 & 113.5397601 & Largo do Senado \\
				12 & 22.19407548 & 113.5394111 & Templo de Sam Kai Vui Kun \\
				13 & 22.19374817 & 113.5402032 & Santa Casa da Miseric\'{o}rdia de Macau \\
				14 & 22.19358846 & 113.5413864 & Igreja da S\'{e} Catedral \\
				15 & 22.19422634 & 113.5412314 & Casa de Lou Kau \\
				16 & 22.19473103 & 113.540409 & Igreja de S\~{a}o Domingos \\
				17 & 22.19788781 & 113.5408688 & Ru\'{i}nas de S\~{a}o Paulo \\
				18 & 22.19774218 & 113.540654 & Templo de Na Tcha \\
				19 & 22.19772073 & 113.5405832 & Old Macau City Walls Sections \\
				20 & 22.19712896 & 113.5422227 & Monte do Forte \\
				21 & 22.19901644 & 113.5393853 & Igreja de Santo Ant\^{o}nio de Lisboa \\
				22 & 22.20036032 & 113.5398153 & Funda\c{c}\~{a}o Oriente \\
				23 & 22.1997325 & 113.5397974 & Cemit\'{e}rio Protestante \\
				24 & 22.19650437 & 113.5496706 & Farol e Fortaleza da Guia \\
				\hline\hline
			\end{tabular}
			\caption{Site ID, latitude, longitude, and names of the Macao World Heritage Sites. Data retrieved from GOOGLE MAP.}
			\label{tab:site data}
		\end{table}
	\end{center}
\begin{sidewaystable}
	\section{Car Driving time for Pairs of Sites}
	\tabcolsep=0.18cm
	\centering
\begin{tabular}{@{}llllllllllllllllllllllllll@{}}
	\toprule
	& 0 & 1 & 2 & 3 & 4 & 5 & 6 & 7 & 8 & 9 & 10 & 11 & 12 & 13 & 14 & 15 & 16 & 17 & 18 & 19 & 20 & 21 & 22 & 23 & 24 \\ \midrule
	\multicolumn{1}{l|}{0} & 0 & 7 & 6 & 6 & 7 & 6 & 8 & 9 & 8 & 9 & 8 & 8 & 8 & 8 & 8 & 10 & 10 & 13 & 13 & 13 & 14 & 11 & 15 & 12 & 11 \\
	\multicolumn{1}{l|}{1} & 4 & 0 & 7 & 7 & 6 & 7 & 9 & 9 & 8 & 9 & 9 & 9 & 10 & 10 & 9 & 11 & 12 & 14 & 14 & 14 & 14 & 11 & 15 & 12 & 11 \\
	\multicolumn{1}{l|}{2} & 4 & 2 & 0 & 1 & 4 & 4 & 6 & 6 & 6 & 5 & 5 & 5 & 6 & 6 & 5 & 6 & 6 & 9 & 9 & 9 & 10 & 8 & 11 & 8 & 9 \\
	\multicolumn{1}{l|}{3} & 5 & 1 & 1 & 0 & 9 & 9 & 10 & 11 & 10 & 10 & 9 & 9 & 10 & 9 & 9 & 11 & 11 & 14 & 14 & 14 & 14 & 12 & 17 & 13 & 11 \\
	\multicolumn{1}{l|}{4} & 5 & 3 & 2 & 1 & 0 & 1 & 3 & 4 & 3 & 4 & 4 & 4 & 5 & 5 & 4 & 6 & 6 & 9 & 9 & 9 & 9 & 6 & 9 & 7 & 7 \\
	\multicolumn{1}{l|}{5} & 9 & 6 & 5 & 5 & 4 & 0 & 2 & 2 & 2 & 2 & 5 & 5 & 5 & 5 & 5 & 7 & 7 & 9 & 9 & 9 & 10 & 6 & 9 & 6 & 8 \\
	\multicolumn{1}{l|}{6} & 7 & 4 & 3 & 3 & 2 & 1 & 0 & 1 & 3 & 1 & 3 & 3 & 5 & 4 & 3 & 5 & 5 & 8 & 8 & 8 & 8 & 6 & 9 & 7 & 7 \\
	\multicolumn{1}{l|}{7} & 6 & 4 & 3 & 3 & 2 & 1 & 3 & 0 & 3 & 1 & 3 & 3 & 5 & 4 & 4 & 5 & 5 & 8 & 8 & 8 & 9 & 6 & 9 & 7 & 7 \\
	\multicolumn{1}{l|}{8} & 7 & 5 & 4 & 4 & 2 & 2 & 1 & 1 & 0 & 1 & 4 & 4 & 5 & 4 & 4 & 6 & 6 & 10 & 11 & 11 & 12 & 6 & 8 & 7 & 8 \\
	\multicolumn{1}{l|}{9} & 6 & 4 & 3 & 3 & 2 & 1 & 3 & 4 & 3 & 0 & 3 & 3 & 5 & 4 & 3 & 5 & 5 & 8 & 8 & 8 & 8 & 6 & 9 & 7 & 6 \\
	\multicolumn{1}{l|}{10} & 9 & 8 & 7 & 8 & 7 & 7 & 5 & 5 & 5 & 5 & 0 & 1 & 1 & 3 & 2 & 5 & 5 & 9 & 9 & 9 & 9 & 4 & 8 & 4 & 5 \\
	\multicolumn{1}{l|}{11} & 9 & 8 & 7 & 7 & 7 & 7 & 6 & 6 & 6 & 6 & 1 & 0 & 2 & 3 & 2 & 5 & 5 & 7 & 7 & 7 & 7 & 4 & 7 & 5 & 6 \\
	\multicolumn{1}{l|}{12} & 11 & 10 & 9 & 9 & 7 & 8 & 6 & 7 & 6 & 6 & 4 & 4 & 0 & 8 & 7 & 10 & 10 & 9 & 9 & 9 & 10 & 3 & 7 & 4 & 11 \\
	\multicolumn{1}{l|}{13} & 16 & 15 & 14 & 14 & 13 & 13 & 14 & 14 & 14 & 14 & 12 & 12 & 14 & 0 & 1 & 4 & 4 & 9 & 9 & 9 & 9 & 12 & 13 & 12 & 7 \\
	\multicolumn{1}{l|}{14} & 16 & 15 & 13 & 13 & 13 & 12 & 14 & 14 & 14 & 14 & 11 & 11 & 12 & 1 & 0 & 3 & 3 & 8 & 7 & 7 & 8 & 11 & 11 & 11 & 6 \\
	\multicolumn{1}{l|}{15} & 16 & 14 & 13 & 13 & 12 & 12 & 13 & 13 & 13 & 13 & 11 & 11 & 12 & 11 & 10 & 0 & 1 & 8 & 8 & 8 & 8 & 11 & 12 & 11 & 7 \\
	\multicolumn{1}{l|}{16} & 16 & 14 & 13 & 13 & 12 & 12 & 13 & 14 & 13 & 13 & 10 & 10 & 12 & 11 & 11 & 1 & 0 & 8 & 8 & 8 & 9 & 11 & 13 & 12 & 7 \\
	\multicolumn{1}{l|}{17} & 20 & 19 & 18 & 18 & 16 & 16 & 15 & 15 & 15 & 15 & 15 & 15 & 15 & 16 & 15 & 11 & 11 & 0 & 1 & 1 & 10 & 4 & 11 & 4 & 12 \\
	\multicolumn{1}{l|}{18} & 19 & 18 & 18 & 17 & 16 & 16 & 15 & 15 & 15 & 15 & 15 & 15 & 15 & 15 & 15 & 11 & 11 & 1 & 0 & 1 & 9 & 4 & 10 & 4 & 12 \\
	\multicolumn{1}{l|}{19} & 23 & 21 & 20 & 20 & 18 & 16 & 16 & 17 & 16 & 16 & 15 & 15 & 16 & 15 & 15 & 11 & 11 & 1 & 1 & 0 & 9 & 4 & 10 & 5 & 12 \\
	\multicolumn{1}{l|}{20} & 14 & 12 & 12 & 11 & 10 & 10 & 11 & 12 & 11 & 12 & 9 & 9 & 10 & 9 & 9 & 2 & 2 & 7 & 7 & 7 & 0 & 10 & 11 & 10 & 5 \\
	\multicolumn{1}{l|}{21} & 18 & 16 & 16 & 15 & 14 & 14 & 13 & 13 & 13 & 13 & 13 & 13 & 13 & 12 & 11 & 8 & 8 & 6 & 6 & 6 & 6 & 0 & 7 & 1 & 8 \\
	\multicolumn{1}{l|}{22} & 18 & 16 & 15 & 15 & 13 & 13 & 11 & 12 & 11 & 11 & 11 & 11 & 11 & 13 & 13 & 8 & 8 & 6 & 6 & 6 & 6 & 9 & 0 & 10 & 9 \\
	\multicolumn{1}{l|}{23} & 19 & 18 & 17 & 17 & 15 & 15 & 17 & 17 & 16 & 15 & 13 & 13 & 15 & 13 & 13 & 9 & 9 & 6 & 6 & 6 & 6 & 9 & 8 & 0 & 9 \\
	\multicolumn{1}{l|}{24} & 12 & 11 & 10 & 9 & 9 & 9 & 11 & 11 & 10 & 11 & 8 & 8 & 9 & 8 & 7 & 9 & 9 & 7 & 7 & 7 & 7 & 10 & 12 & 11 & 0 \\ \bottomrule
\end{tabular}
	\caption{Car driving time (in units of minutes) for pairs of sites. The Site ID number is given in Table A.\ref{tab:site data}. Site ID numbers at the left-hand side column mean the departure site, while Site ID numbers at the top row refer to the destination site.}
	\label{tab:car time}
\end{sidewaystable}
\begin{sidewaystable}
	\section{Car Driving distance for Pairs of Sites }
	\tabcolsep=0.13cm
	\centering     
	\begin{tabular}{@{}llllllllllllllllllllllllll@{}}
		\toprule
		& 0 & 1 & 2 & 3 & 4 & 5 & 6 & 7 & 8 & 9 & 10 & 11 & 12 & 13 & 14 & 15 & 16 & 17 & 18 & 19 & 20 & 21 & 22 & 23 & 24 \\ \midrule
		\multicolumn{1}{l|}{0} & 0 & 1.9 & 1.6 & 1.6 & 1.8 & 1.8 & 2.2 & 2.2 & 2.1 & 2.2 & 2.1 & 2.1 & 2.3 & 2.3 & 2.1 & 2.6 & 2.6 & 3.4 & 3.4 & 3.5 & 3.5 & 3.5 & 3.7 & 3.6 & 4.7 \\
		\multicolumn{1}{l|}{1} & 1.3 & 0 & 2.5 & 2.2 & 2.8 & 2.7 & 3.5 & 3.5 & 3.1 & 3 & 3.3 & 3.3 & 3.5 & 3.5 & 3.3 & 3.8 & 3.8 & 4.7 & 4.7 & 4.8 & 4.7 & 4.7 & 4.5 & 4.4 & 5.1 \\
		\multicolumn{1}{l|}{2} & 1.2 & 0.3 & 0 & 0.65 & 0.9 & 0.9 & 1.3 & 1.3 & 1.2 & 1.2 & 1.2 & 1.2 & 1.4 & 1.4 & 3.1 & 3.6 & 3.6 & 4.9 & 4.9 & 4.9 & 4.8 & 4.8 & 2.8 & 2.2 & 2.4 \\
		\multicolumn{1}{l|}{3} & 1.6 & 0.3 & 0.065 & 0 & 3.1 & 3 & 3.4 & 3.4 & 3.4 & 3.3 & 3.3 & 3.6 & 3.8 & 3.8 & 3.8 & 4.1 & 4.1 & 5 & 5 & 5.1 & 5 & 5 & 4.8 & 2.2 & 4.8 \\
		\multicolumn{1}{l|}{4} & 1.5 & 0.6 & 0.35 & 0.065 & 0 & 0.3 & 0.7 & 0.7 & 0.6 & 0.7 & 0.85 & 0.85 & 1.05 & 1.05 & 0.85 & 1.4 & 1.4 & 2.2 & 2.2 & 2.3 & 2.3 & 2.3 & 2.2 & 1.6 & 2.1 \\
		\multicolumn{1}{l|}{5} & 2.2 & 1.4 & 1.1 & 1.1 & 0.8 & 0 & 0.4 & 0.4 & 0.3 & 0.4 & 0.55 & 0.55 & 1.2 & 1.2 & 1.1 & 1.6 & 1.6 & 2.3 & 2.3 & 2.4 & 2.4 & 2.4 & 1.9 & 1.3 & 2.3 \\
		\multicolumn{1}{l|}{6} & 1.9 & 1 & 0.75 & 0.75 & 0.4 & 0.4 & 0 & 0.026 & 0.7 & 0.32 & 0.65 & 0.65 & 0.85 & 0.85 & 0.7 & 1.2 & 1.2 & 2 & 2 & 2.1 & 2.1 & 1.6 & 2.3 & 1.7 & 1.9 \\
		\multicolumn{1}{l|}{7} & 1.8 & 1 & 0.75 & 0.75 & 0.4 & 0.4 & 0.75 & 0 & 0.7 & 0.32 & 0.65 & 0.65 & 0.85 & 0.85 & 0.7 & 1.2 & 1.2 & 2 & 2 & 2.1 & 2.1 & 1.6 & 2.3 & 1.7 & 1.9 \\
		\multicolumn{1}{l|}{8} & 2 & 1.1 & 0.85 & 0.7 & 0.5 & 0.5 & 0.11 & 0.14 & 0 & 0.14 & 0.5 & 0.75 & 1 & 1 & 0.8 & 1.3 & 1.3 & 2.1 & 2.1 & 2.2 & 2.2 & 1.1 & 1.7 & 1.2 & 2 \\
		\multicolumn{1}{l|}{9} & 1.8 & 1 & 0.7 & 0.55 & 0.35 & 0.35 & 0.7 & 0.75 & 0.65 & 0 & 0.75 & 1 & 1.25 & 1.25 & 1.05 & 1.2 & 1.2 & 2 & 2 & 2.1 & 2.1 & 1.5 & 2.2 & 1.6 & 1.8 \\
		\multicolumn{1}{l|}{10} & 2.4 & 1.6 & 1.4 & 1.4 & 1.3 & 1.3 & 1.1 & 1.2 & 1.1 & 1.2 & 0 & 0.005 & 0.21 & 0.21 & 0.4 & 0.85 & 0.85 & 1.8 & 1.8 & 1.9 & 1.9 & 0.9 & 1.7 & 1 & 1.6 \\
		\multicolumn{1}{l|}{11} & 2.4 & 1.6 & 1.4 & 1.4 & 1.3 & 1.3 & 1.1 & 1.2 & 1.1 & 1.2 & 1.2 & 0 & 0.21 & 0.21 & 0.4 & 0.9 & 0.9 & 1.8 & 1.8 & 1.9 & 1.9 & 0.9 & 1.7 & 1 & 1.6 \\
		\multicolumn{1}{l|}{12} & 3 & 2.3 & 2 & 2 & 1.7 & 1.7 & 1.5 & 1.5 & 1.4 & 1.5 & 1.5 & 0.8 & 0 & 1.3 & 0.4 & 1.7 & 1.7 & 1.9 & 1.9 & 2 & 2 & 1 & 1.4 & 0.9 & 2.4 \\
		\multicolumn{1}{l|}{13} & 3.7 & 3 & 2.8 & 2.8 & 2.4 & 2.4 & 2.8 & 2.8 & 2.7 & 2.8 & 2.8 & 2 & 2.2 & 0 & 0.16 & 0.65 & 0.65 & 1.5 & 1.5 & 1.6 & 1.6 & 2.1 & 2.5 & 2.2 & 1.4 \\
		\multicolumn{1}{l|}{14} & 3.6 & 2.9 & 2.6 & 2.6 & 2.3 & 2.3 & 2.6 & 2.6 & 2.5 & 2.6 & 2.6 & 1.8 & 2 & 0.16 & 0 & 0.5 & 0.5 & 1.4 & 1.4 & 1.4 & 1.4 & 1.9 & 2.4 & 2 & 1.2 \\
		\multicolumn{1}{l|}{15} & 3.9 & 3.2 & 2.9 & 2.9 & 2.6 & 2.6 & 2.9 & 2.9 & 2.8 & 2.9 & 2.9 & 2.1 & 2.3 & 0.36 & 2.2 & 0 & 0.68 & 1.7 & 1.7 & 1.8 & 1.8 & 2.3 & 2.7 & 2.3 & 1.5 \\
		\multicolumn{1}{l|}{16} & 3.9 & 3.2 & 2.9 & 2.9 & 2.6 & 2.6 & 2.9 & 2.9 & 2.8 & 2.9 & 2.9 & 2.1 & 2.3 & 0.36 & 2.2 & 0.068 & 0 & 1.7 & 1.7 & 1.8 & 1.8 & 2.3 & 2.7 & 2.3 & 1.5 \\
		\multicolumn{1}{l|}{17} & 4.8 & 4.1 & 3.8 & 3.8 & 3.6 & 3.6 & 3.4 & 3.4 & 3.3 & 3.4 & 3.4 & 3.6 & 3.1 & 3.3 & 3.1 & 2.2 & 2.2 & 0 & 0 & 0.007 & 1.8 & 0.55 & 2.1 & 0.7 & 2.4 \\
		\multicolumn{1}{l|}{18} & 4.7 & 4 & 3.7 & 3.7 & 3.4 & 3.4 & 3.3 & 3.3 & 3.2 & 3.3 & 3.3 & 3.5 & 3 & 3.2 & 3 & 2.1 & 2.1 & 0 & 0 & 0.007 & 1.8 & 0.55 & 2.1 & 0.7 & 2.4 \\
		\multicolumn{1}{l|}{19} & 4.8 & 4.1 & 4 & 4 & 3.5 & 3.5 & 3.4 & 3.4 & 3.3 & 3.4 & 3.4 & 2.6 & 3.1 & 3.3 & 3.1 & 2.2 & 2.2 & 0.1 & 0.1 & 0 & 1.8 & 0.6 & 2.1 & 0.7 & 2.4 \\
		\multicolumn{1}{l|}{20} & 3.5 & 2.8 & 2.6 & 2.6 & 2.3 & 2.3 & 2.2 & 2.2 & 2.5 & 2.2 & 2.2 & 1.4 & 2 & 2.1 & 1.9 & 0.45 & 0.45 & 1.3 & 1.3 & 1.4 & 0 & 1.9 & 2.4 & 2 & 1.2 \\
		\multicolumn{1}{l|}{21} & 4.2 & 3.5 & 3.4 & 3.4 & 2.9 & 2.9 & 2.8 & 2.8 & 2.7 & 2.8 & 2.8 & 2 & 2.6 & 2.7 & 2.6 & 1.7 & 1.7 & 1.1 & 1.1 & 1.2 & 1.2 & 0 & 1.5 & 0.12 & 1.8 \\
		\multicolumn{1}{l|}{22} & 4.2 & 3.4 & 3.1 & 3.1 & 2.8 & 2.8 & 2.6 & 2.6 & 2.5 & 2.6 & 2.6 & 1.8 & 2.4 & 2.5 & 2.5 & 1.7 & 1.7 & 1.1 & 1.1 & 1.2 & 1.2 & 1.7 & 0 & 1.8 & 1.8 \\
		\multicolumn{1}{l|}{23} & 4.3 & 3 & 3.4 & 3.4 & 3 & 3 & 2.9 & 2.9 & 2.8 & 2.9 & 2.9 & 2.1 & 2.7 & 2.8 & 2.6 & 1.7 & 1.7 & 1.2 & 1.2 & 1.2 & 1.2 & 1.7 & 1.6 & 0 & 1.9 \\
		\multicolumn{1}{l|}{24} & 4.3 & 3.1 & 3.5 & 3.5 & 2.1 & 2.1 & 2.4 & 2.4 & 2.3 & 2.4 & 2.4 & 1.6 & 2.2 & 2.3 & 1.7 & 1.8 & 1.8 & 1.5 & 1.5 & 1.6 & 1.6 & 2 & 2.5 & 2.1 & 0 \\
		\bottomrule
	\end{tabular}
	\caption{Car driving distance (in units of km) for pairs of sites. The Site ID number is given in Table A.\ref{tab:site data}. Site ID numbers at the left-hand side column mean the departure site, while Site ID numbers at the top row refer to the destination site.}
	\label{tab:car distance}
\end{sidewaystable}
\begin{sidewaystable}
	\section{Bus Driving time for Pairs of Sites }
	\tabcolsep=0.16cm
	\centering
	\begin{tabular}{@{}llllllllllllllllllllllllll@{}}
		\toprule
		& 0 & 1 & 2 & 3 & 4 & 5 & 6 & 7 & 8 & 9 & 10 & 11 & 12 & 13 & 14 & 15 & 16 & 17 & 18 & 19 & 20 & 21 & 22 & 23 & 24 \\ \midrule
		\multicolumn{1}{l|}{0} & 0 & 4$\star$ & 34 & 8$\star$ & 23 & 25 & 25 & 24 & 27 & 24 & 25 & 26 & 26 & 26 & 27 & 28 & 30 & 31 & 31 & 35 & 35 & 30 & 34 & 32 & 39 \\
		\multicolumn{1}{l|}{1} & 4$\star$ & 0 & 4$\star$ & 4$\star$ & 31 & 30 & 27 & 26 & 28 & 26 & 26 & 26 & 26 & 27 & 21 & 22 & 23 & 28 & 28 & 30 & 27 & 31 & 32 & 31 & 36 \\
		\multicolumn{1}{l|}{2} & 12 & 4$\star$ & 0 & 1$\star$ & 4$\star$ & 7$\star$ & 28 & 27 & 30 & 27 & 23 & 23 & 24 & 24 & 17 & 23 & 24 & 28 & 28 & 37 & 28 & 34 & 36 & 35 & 32 \\
		\multicolumn{1}{l|}{3} & 12 & 4$\star$ & 1$\star$ & 0 & 4$\star$ & 6$\star$ & 28 & 27 & 29 & 27 & 22 & 22 & 24 & 24 & 22 & 23 & 24 & 28 & 28 & 36 & 27 & 34 & 36 & 35 & 31 \\
		\multicolumn{1}{l|}{4} & 14 & 13 & 10 & 10 & 0 & 2$\star$ & 4$\star$ & 5$\star$ & 6$\star$ & 5$\star$ & 22 & 23 & 24 & 23 & 21 & 23 & 24 & 28 & 28 & 31 & 27 & 22 & 25 & 23 & 30 \\
		\multicolumn{1}{l|}{5} & 16 & 15 & 13 & 12 & 2$\star$ & 0 & 5$\star$ & 4$\star$ & 4$\star$ & 4$\star$ & 6$\star$ & 6$\star$ & 6$\star$ & 7$\star$ & 23 & 25 & 9$\star$ & 11$\star$ & 11$\star$ & 35 & 27 & 20 & 22 & 20 & 32 \\
		\multicolumn{1}{l|}{6} & 19 & 18 & 18 & 18 & 4$\star$ & 5$\star$ & 0 & 1$\star$ & 1$\star$ & 1$\star$ & 4$\star$ & 4$\star$ & 4$\star$ & 4$\star$ & 6$\star$ & 6$\star$ & 6$\star$ & 10$\star$ & 10$\star$ & 25 & 21 & 17 & 19 & 17 & 28 \\
		\multicolumn{1}{l|}{7} & 18 & 16 & 17 & 17 & 5$\star$ & 4$\star$ & 1$\star$ & 0 & 2$\star$ & 1$\star$ & 4$\star$ & 4$\star$ & 5$\star$ & 5$\star$ & 6$\star$ & 7$\star$ & 7$\star$ & 11$\star$ & 11$\star$ & 26 & 20 & 18 & 20 & 19 & 28 \\
		\multicolumn{1}{l|}{8} & 18 & 19 & 20 & 19 & 6$\star$ & 4$\star$ & 1$\star$ & 2$\star$ & 0 & 2$\star$ & 3$\star$ & 4$\star$ & 4$\star$ & 5$\star$ & 6$\star$ & 6$\star$ & 6$\star$ & 10$\star$ & 10$\star$ & 13$\star$ & 23 & 18 & 19 & 18 & 31 \\
		\multicolumn{1}{l|}{9} & 18 & 16 & 17 & 17 & 5$\star$ & 4$\star$ & 1$\star$ & 1$\star$ & 2$\star$ & 0 & 4$\star$ & 4$\star$ & 5$\star$ & 5$\star$ & 6$\star$ & 7$\star$ & 7$\star$ & 11$\star$ & 11$\star$ & 26 & 21 & 18 & 20 & 19 & 29 \\
		\multicolumn{1}{l|}{10} & 19 & 21 & 21 & 20 & 25 & 6$\star$ & 4$\star$ & 4$\star$ & 3$\star$ & 4$\star$ & 0 & 1$\star$ & 1$\star$ & 1$\star$ & 3$\star$ & 3$\star$ & 3$\star$ & 7$\star$ & 7$\star$ & 10$\star$ & 11$\star$ & 15 & 17 & 16 & 28 \\
		\multicolumn{1}{l|}{11} & 19 & 21 & 20 & 20 & 24 & 6$\star$ & 4$\star$ & 4$\star$ & 4$\star$ & 4$\star$ & 1$\star$ & 0 & 1$\star$ & 1$\star$ & 3$\star$ & 3$\star$ & 3$\star$ & 7$\star$ & 7$\star$ & 10$\star$ & 11$\star$ & 16 & 16 & 18 & 27 \\
		\multicolumn{1}{l|}{12} & 19 & 20 & 19 & 19 & 18 & 6$\star$ & 5$\star$ & 5$\star$ & 5$\star$ & 5$\star$ & 1$\star$ & 1$\star$ & 0 & 2$\star$ & 3$\star$ & 3$\star$ & 3$\star$ & 6$\star$ & 6$\star$ & 9$\star$ & 10$\star$ & 14 & 17 & 15 & 27 \\
		\multicolumn{1}{l|}{13} & 21 & 22 & 22 & 21 & 21 & 8$\star$ & 6$\star$ & 6$\star$ & 6$\star$ & 6$\star$ & 2$\star$ & 2$\star$ & 2$\star$ & 0 & 2$\star$ & 2$\star$ & 2$\star$ & 6$\star$ & 6$\star$ & 9$\star$ & 10$\star$ & 15 & 18 & 16 & 27 \\
		\multicolumn{1}{l|}{14} & 21 & 20 & 18 & 18 & 18 & 21 & 7$\star$ & 7$\star$ & 7$\star$ & 7$\star$ & 3$\star$ & 3$\star$ & 3$\star$ & 2$\star$ & 0 & 1$\star$ & 2$\star$ & 6$\star$ & 6$\star$ & 9$\star$ & 8$\star$ & 16 & 18 & 17 & 25 \\
		\multicolumn{1}{l|}{15} & 22 & 22 & 21 & 21 & 20 & 10$\star$ & 7$\star$ & 7$\star$ & 7$\star$ & 8$\star$ & 3$\star$ & 3$\star$ & 3$\star$ & 2$\star$ & 1$\star$ & 0 & 1$\star$ & 5$\star$ & 5$\star$ & 8$\star$ & 8$\star$ & 15 & 17 & 16 & 27 \\
		\multicolumn{1}{l|}{16} & 23 & 23 & 23 & 21 & 21 & 10$\star$ & 7$\star$ & 8$\star$ & 7$\star$ & 8$\star$ & 4$\star$ & 4$\star$ & 3$\star$ & 2$\star$ & 3$\star$ & 1$\star$ & 0 & 5$\star$ & 5$\star$ & 7$\star$ & 7$\star$ & 15 & 18 & 16 & 28 \\
		\multicolumn{1}{l|}{17} & 22 & 23 & 23 & 24 & 20 & 12$\star$ & 11$\star$ & 11$\star$ & 11$\star$ & 11$\star$ & 7$\star$ & 7$\star$ & 6$\star$ & 6$\star$ & 6$\star$ & 5$\star$ & 4$\star$ & 0 & 1$\star$ & 2$\star$ & 10$\star$ & 3$\star$ & 6$\star$ & 5$\star$ & 32 \\
		\multicolumn{1}{l|}{18} & 22 & 23 & 23 & 23 & 20 & 11$\star$ & 11$\star$ & 11$\star$ & 10$\star$ & 11$\star$ & 7$\star$ & 7$\star$ & 6$\star$ & 5$\star$ & 6$\star$ & 4$\star$ & 4$\star$ & 1$\star$ & 0 & 1$\star$ & 10$\star$ & 3$\star$ & 6$\star$ & 5$\star$ & 32 \\
		\multicolumn{1}{l|}{19} & 25 & 26 & 26 & 26 & 23 & 14$\star$ & 14$\star$ & 14$\star$ & 13$\star$ & 14$\star$ & 10$\star$ & 10$\star$ & 9$\star$ & 8$\star$ & 9$\star$ & 7$\star$ & 7$\star$ & 2$\star$ & 1$\star$ & 0 & 10$\star$ & 6$\star$ & 9$\star$ & 8$\star$ & 38 \\
		\multicolumn{1}{l|}{20} & 26 & 25 & 23 & 23 & 23 & 26 & 13$\star$ & 13$\star$ & 13$\star$ & 23 & 9$\star$ & 9$\star$ & 8$\star$ & 8$\star$ & 6$\star$ & 6$\star$ & 5$\star$ & 9$\star$ & 9$\star$ & 9$\star$ & 0 & 20 & 22 & 20 & 29 \\
		\multicolumn{1}{l|}{21} & 21 & 22 & 22 & 21 & 18 & 17 & 16 & 16 & 16 & 17 & 12 & 12 & 8$\star$ & 9$\star$ & 9$\star$ & 8$\star$ & 7$\star$ & 4$\star$ & 4$\star$ & 7$\star$ & 13$\star$ & 0 & 2$\star$ & 1$\star$ & 31 \\
		\multicolumn{1}{l|}{22} & 25 & 24 & 24 & 23 & 21 & 21 & 20 & 20 & 19 & 21 & 18 & 17 & 16 & 19 & 22 & 10$\star$ & 9$\star$ & 6$\star$ & 6$\star$ & 9$\star$ & 14$\star$ & 2$\star$ & 0 & 1$\star$ & 32 \\
		\multicolumn{1}{l|}{23} & 24 & 23 & 23 & 22 & 19 & 19 & 19 & 18 & 18 & 19 & 15 & 15 & 9$\star$ & 10$\star$ & 11$\star$ & 9$\star$ & 8$\star$ & 5$\star$ & 5$\star$ & 8$\star$ & 13$\star$ & 1$\star$ & 1$\star$ & 0 & 30 \\
		\multicolumn{1}{l|}{24} & 32 & 31 & 31 & 31 & 29 & 30 & 29 & 29 & 29 & 30 & 26 & 25 & 25 & 26 & 25 & 25 & 25 & 29 & 29 & 18$\star$ & 26 & 26 & 27 & 26 & 0 \\ \bottomrule
	\end{tabular}
	\caption{Bus driving time (in units of minutes) for pairs of sites. If, for a pair of sites, there is no suggested route for bus that may be found from the google map, then the value is replaced with the least time for the pedestrians to walk across the two sites, indicated with $\star$. The Site ID number is given in Table A.\ref{tab:site data}. Site ID numbers at the left-hand side column mean the departure site, while Site ID numbers at the top row refer to the destination site.}
	\label{tab:bus time}
\end{sidewaystable}
\begin{sidewaystable}
	\section{Pedestrian walking time for Pairs of Sites }
	\tabcolsep=0.18cm
	\centering
	\begin{tabular}{@{}llllllllllllllllllllllllll@{}}
		\toprule
		& 0 & 1 & 2 & 3 & 4 & 5 & 6 & 7 & 8 & 9 & 10 & 11 & 12 & 13 & 14 & 15 & 16 & 17 & 18 & 19 & 20 & 21 & 22 & 23 & 24 \\ \midrule
		\multicolumn{1}{l|}{0} & 0 & 4 & 8 & 8 & 12 & 13 & 16 & 16 & 16 & 14 & 18 & 18 & 19 & 19 & 18 & 20 & 21 & 24 & 24 & 27 & 25 & 23 & 26 & 24 & 35 \\
		\multicolumn{1}{l|}{1} & 4 & 0 & 4 & 4 & 8 & 10 & 12 & 12 & 13 & 11 & 15 & 15 & 16 & 16 & 15 & 17 & 18 & 22 & 22 & 25 & 22 & 24 & 26 & 25 & 32 \\
		\multicolumn{1}{l|}{2} & 8 & 4 & 0 & 1 & 4 & 7 & 9 & 9 & 10 & 8 & 12 & 12 & 13 & 13 & 13 & 14 & 16 & 19 & 19 & 22 & 19 & 21 & 23 & 22 & 30 \\
		\multicolumn{1}{l|}{3} & 8 & 4 & 1 & 0 & 4 & 6 & 9 & 8 & 10 & 8 & 12 & 12 & 13 & 13 & 12 & 13 & 15 & 19 & 19 & 22 & 19 & 20 & 22 & 21 & 29 \\
		\multicolumn{1}{l|}{4} & 12 & 8 & 4 & 4 & 0 & 3 & 6 & 5 & 7 & 4 & 8 & 9 & 9 & 10 & 9 & 10 & 12 & 15 & 14 & 17 & 16 & 17 & 19 & 18 & 26 \\
		\multicolumn{1}{l|}{5} & 13 & 10 & 7 & 6 & 3 & 0 & 5 & 5 & 4 & 4 & 7 & 7 & 7 & 9 & 9 & 10 & 10 & 12 & 12 & 15 & 14 & 14 & 17 & 16 & 25 \\
		\multicolumn{1}{l|}{6} & 16 & 12 & 9 & 9 & 6 & 5 & 0 & 1 & 2 & 4 & 4 & 4 & 5 & 6 & 7 & 8 & 8 & 11 & 11 & 14 & 13 & 14 & 16 & 15 & 23 \\
		\multicolumn{1}{l|}{7} & 16 & 12 & 9 & 8 & 5 & 5 & 1 & 0 & 2 & 1 & 5 & 4 & 6 & 7 & 7 & 8 & 8 & 12 & 11 & 14 & 14 & 14 & 17 & 16 & 24 \\
		\multicolumn{1}{l|}{8} & 16 & 13 & 10 & 10 & 7 & 4 & 2 & 2 & 0 & 2 & 4 & 4 & 5 & 6 & 7 & 8 & 8 & 11 & 11 & 14 & 13 & 13 & 16 & 15 & 24 \\
		\multicolumn{1}{l|}{9} & 14 & 11 & 8 & 8 & 4 & 4 & 1 & 1 & 2 & 0 & 5 & 5 & 6 & 7 & 7 & 8 & 8 & 12 & 12 & 14 & 14 & 14 & 17 & 16 & 24 \\
		\multicolumn{1}{l|}{10} & 18 & 15 & 12 & 12 & 8 & 7 & 4 & 5 & 4 & 5 & 0 & 1 & 2 & 2 & 3 & 4 & 4 & 8 & 7 & 10 & 9 & 10 & 15 & 11 & 17 \\
		\multicolumn{1}{l|}{11} & 18 & 15 & 12 & 12 & 9 & 7 & 4 & 4 & 4 & 5 & 1 & 0 & 1 & 2 & 3 & 4 & 4 & 7 & 7 & 10 & 9 & 10 & 13 & 11 & 20 \\
		\multicolumn{1}{l|}{12} & 19 & 16 & 13 & 13 & 9 & 7 & 5 & 6 & 5 & 6 & 2 & 1 & 0 & 3 & 4 & 3 & 3 & 6 & 6 & 9 & 8 & 9 & 11 & 10 & 19 \\
		\multicolumn{1}{l|}{13} & 19 & 16 & 13 & 13 & 10 & 9 & 6 & 7 & 6 & 7 & 2 & 2 & 3 & 0 & 2 & 2 & 3 & 6 & 6 & 9 & 8 & 9 & 12 & 10 & 18 \\
		\multicolumn{1}{l|}{14} & 18 & 15 & 13 & 12 & 9 & 9 & 7 & 7 & 7 & 7 & 3 & 3 & 4 & 2 & 0 & 2 & 3 & 7 & 7 & 9 & 7 & 10 & 12 & 11 & 17 \\
		\multicolumn{1}{l|}{15} & 20 & 17 & 14 & 13 & 10 & 10 & 8 & 8 & 8 & 8 & 4 & 4 & 3 & 2 & 2 & 0 & 2 & 5 & 4 & 8 & 6 & 8 & 11 & 9 & 17 \\
		\multicolumn{1}{l|}{16} & 21 & 18 & 16 & 15 & 12 & 10 & 8 & 8 & 8 & 8 & 4 & 4 & 3 & 3 & 3 & 2 & 0 & 4 & 5 & 7 & 6 & 8 & 10 & 9 & 17 \\
		\multicolumn{1}{l|}{17} & 24 & 22 & 19 & 19 & 15 & 12 & 11 & 12 & 11 & 12 & 8 & 7 & 6 & 6 & 7 & 5 & 4 & 0 & 1 & 3 & 9 & 5 & 7 & 6 & 19 \\
		\multicolumn{1}{l|}{18} & 24 & 22 & 19 & 19 & 14 & 12 & 11 & 11 & 11 & 12 & 7 & 7 & 6 & 6 & 7 & 5 & 4 & 1 & 0 & 1 & 9 & 7 & 7 & 6 & 19 \\
		\multicolumn{1}{l|}{19} & 27 & 25 & 22 & 22 & 17 & 15 & 14 & 14 & 14 & 14 & 10 & 10 & 9 & 9 & 9 & 8 & 7 & 3 & 1 & 0 & 9 & 7 & 10 & 9 & 19 \\
		\multicolumn{1}{l|}{20} & 25 & 22 & 19 & 19 & 16 & 14 & 13 & 14 & 13 & 14 & 9 & 9 & 8 & 8 & 7 & 6 & 6 & 9 & 9 & 9 & 0 & 13 & 15 & 13 & 17 \\
		\multicolumn{1}{l|}{21} & 23 & 24 & 21 & 20 & 17 & 14 & 14 & 14 & 13 & 14 & 10 & 10 & 9 & 9 & 10 & 8 & 8 & 5 & 7 & 7 & 13 & 0 & 2 & 1 & 21 \\
		\multicolumn{1}{l|}{22} & 26 & 26 & 23 & 22 & 19 & 17 & 16 & 17 & 16 & 17 & 15 & 13 & 11 & 12 & 12 & 11 & 10 & 7 & 7 & 10 & 15 & 2 & 0 & 2 & 22 \\
		\multicolumn{1}{l|}{23} & 24 & 25 & 22 & 21 & 18 & 16 & 15 & 16 & 15 & 16 & 11 & 11 & 10 & 10 & 11 & 9 & 9 & 6 & 6 & 9 & 13 & 1 & 2 & 0 & 21 \\
		\multicolumn{1}{l|}{24} & 35 & 32 & 30 & 29 & 26 & 25 & 23 & 24 & 24 & 24 & 17 & 20 & 19 & 18 & 17 & 17 & 17 & 19 & 19 & 19 & 17 & 21 & 22 & 21 & 0 \\
		\bottomrule
	\end{tabular}
	\caption{Pedestrian walking time (in units of minutes) for pairs of sites. The Site ID number is given in Table A.\ref{tab:site data}. Site ID numbers at the left-hand side column mean the departure site, while Site ID numbers at the top row refer to the destination site.}
	\label{tab:pedestrian time}
\end{sidewaystable}
\begin{sidewaystable}
	\section{Pedestrian walking distance for Pairs of Sites }
	\tabcolsep=0.1cm
	\centering     
	\begin{tabular}{@{}llllllllllllllllllllllllll@{}}
		\toprule
		& 0 & 1 & 2 & 3 & 4 & 5 & 6 & 7 & 8 & 9 & 10 & 11 & 12 & 13 & 14 & 15 & 16 & 17 & 18 & 19 & 20 & 21 & 22 & 23 & 24 \\ \midrule
	\multicolumn{1}{l|}{0} & 0 & 0.24 & 0.55 & 0.55 & 0.85 & 1 & 1.1 & 1.1 & 1.2 & 1.1 & 1.3 & 1.3 & 1.5 & 1.4 & 1.4 & 1.5 & 1.6 & 1.9 & 1.9 & 2.1 & 1.9 & 1.8 & 2 & 1.9 & 2.6 \\
	\multicolumn{1}{l|}{1} & 0.24 & 0 & 0.3 & 0.3 & 0.6 & 0.75 & 0.9 & 0.85 & 1 & 0.85 & 1.1 & 1.1 & 1.2 & 1.2 & 1.2 & 1.2 & 1.4 & 1.7 & 1.6 & 1.8 & 1.7 & 1.8 & 2 & 1.9 & 2.4 \\
	\multicolumn{1}{l|}{2} & 0.55 & 0.3 & 0 & 0.064 & 0.35 & 0.5 & 0.65 & 0.6 & 0.75 & 0.6 & 0.85 & 0.85 & 0.95 & 0.9 & 0.9 & 1 & 1.1 & 1.4 & 1.4 & 1.6 & 1.4 & 1.6 & 1.8 & 1.7 & 2.2 \\
	\multicolumn{1}{l|}{3} & 0.55 & 0.3 & 0.064 & 0 & 0.3 & 0.5 & 0.6 & 0.6 & 0.75 & 0.6 & 0.85 & 0.85 & 0.95 & 0.9 & 0.9 & 0.95 & 1.1 & 1.4 & 1.4 & 1.6 & 1.4 & 1.5 & 1.7 & 1.6 & 2.1 \\
	\multicolumn{1}{l|}{4} & 0.85 & 0.6 & 0.35 & 0.3 & 0 & 0.18 & 0.4 & 0.35 & 0.5 & 0.35 & 0.6 & 0.6 & 0.65 & 0.65 & 0.65 & 0.75 & 0.9 & 1.1 & 1.1 & 1.2 & 1.2 & 1.2 & 1.4 & 1.3 & 1.9 \\
	\multicolumn{1}{l|}{5} & 1 & 0.75 & 0.5 & 0.5 & 0.18 & 0 & 0.4 & 0.35 & 0.3 & 0.35 & 0.45 & 0.45 & 	0.5 & 0.55 & 0.65 & 0.7 & 0.7 & 0.9 & 0.9 & 1.1 & 1 & 1.1 & 1.3 & 1.1 & 1.9 \\
	\multicolumn{1}{l|}{6} & 1.1 & 0.9 & 0.65 & 0.6 & 0.4 & 0.4 & 0 & 0.026 & 0.11 & 0.032 & 0.25 & 0.25 & 0.3 & 0.4 & 0.45 & 0.5 & 0.5 & 0.8 & 0.8 & 1 & 0.85 & 1 & 1 & 1.1 & 1.6 \\
	\multicolumn{1}{l|}{7} & 1.1 & 0.85 & 0.6 & 0.6 & 0.35 & 0.35 & 0.026 & 0 & 0.14 & 0.005 & 0.28 & 0.27 & 0.28 & 0.4 & 0.45 & 0.5 & 0.55 & 0.85 & 0.8 & 1 & 0.9 & 1 & 1.2 & 1.1 & 1.7 \\
	\multicolumn{1}{l|}{8} & 1.2 & 1 & 0.75 & 0.75 & 0.5 & 0.3 & 0.11 & 0.14 & 0 & 0.14 & 0.27 & 0.27 & 0.35 & 0.4 & 0.5 & 0.5 & 0.5 & 0.8 & 0.75 & 1 & 0.9 & 0.95 & 1.1 & 1 & 1.7 \\
	\multicolumn{1}{l|}{9} & 1.1 & 0.85 & 0.6 & 0.6 & 0.35 & 0.35 & 0.032 & 0.005 & 0.14 & 0 & 0.28 & 0.28 & 0.35 & 0.4 & 0.45 & 0.55 & 0.55 & 0.85 & 0.8 & 1 & 0.9 & 1 & 1.2 & 1.1 & 1.7 \\
	\multicolumn{1}{l|}{10} & 1.3 & 1.1 & 0.85 & 0.85 & 0.6 & 0.45 & 0.25 & 0.28 & 0.27 & 0.28 & 0 & 0.005 & 0.11 & 0.12 & 0.22 & 0.27 & 0.3 & 0.6 & 0.6 & 0.8 & 0.65 & 0.8 & 1 & 0.9 & 1.4 \\
	\multicolumn{1}{l|}{11} & 1.3 & 1.1 & 0.85 & 0.85 & 0.6 & 0.45 & 0.25 & 0.27 & 0.27 & 0.28 & 0.005 & 0 & 0.11 & 0.12 & 0.22 & 0.27 & 0.29 & 0.6 & 0.6 & 0.75 & 0.65 & 0.8 & 1 & 0.9 & 1.4 \\
	\multicolumn{1}{l|}{12} & 1.5 & 1.2 & 0.95 & 0.95 & 0.65 & 0.5 & 0.3 & 0.28 & 0.35 & 0.35 & 0.11 & 0.11 & 0 & 0.2 & 0.26 & 0.24 & 0.25 & 0.5 & 0.5 & 0.7 & 0.6 & 0.7 & 0.85 & 0.75 & 1.4 \\
	\multicolumn{1}{l|}{13} & 1.4 & 1.2 & 0.9 & 0.9 & 0.65 & 0.55 & 0.4 & 0.4 & 0.4 & 0.4 & 0.12 & 0.12 & 0.2 & 0 & 0.18 & 0.18 & 0.22 & 0.5 & 0.5 & 0.7 & 0.65 & 0.75 & 0.9 & 0.8 & 1.4 \\
	\multicolumn{1}{l|}{14} & 1.4 & 1.2 & 0.9 & 0.9 & 0.65 & 0.65 & 0.45 & 0.45 & 0.5 & 0.45 & 0.22 & 0.22 & 0.26 & 0.18 & 0 & 0.12 & 0.22 & 0.5 & 0.5 & 0.7 & 0.5 & 0.75 & 0.95 & 0.85 & 1.2 \\
	\multicolumn{1}{l|}{15} & 1.5 & 1.2 & 1 & 0.95 & 0.75 & 0.7 & 0.5 & 0.5 & 0.5 & 0.55 & 0.27 & 0.27 & 0.24 & 0.18 & 0.12 & 0 & 0.12 & 0.4 & 0.4 & 0.6 & 0.5 & 0.65 & 0.85 & 0.75 & 1.3 \\
	\multicolumn{1}{l|}{16} & 1.6 & 1.4 & 1.1 & 1.1 & 0.9 & 0.7 & 0.5 & 0.55 & 0.5 & 0.55 & 0.3 & 0.29 & 0.25 & 0.22 & 0.22 & 0.12 & 0 & 0.4 & 0.35 & 0.55 & 0.45 & 0.65 & 0.85 & 0.7 & 1.3 \\
	\multicolumn{1}{l|}{17} & 1.9 & 1.7 & 1.4 & 1.4 & 1.1 & 0.9 & 0.8 & 0.85 & 0.8 & 0.85 & 0.6 & 0.6 & 0.5 & 0.5 & 0.5 & 0.4 & 0.4 & 0 & 0.008 & 0.23 & 0.65 & 0.35 & 0.5 & 0.4 & 1.3 \\
	\multicolumn{1}{l|}{18} & 1.9 & 1.6 & 1.4 & 1.4 & 1.1 & 0.9 & 0.8 & 0.8 & 0.75 & 0.8 & 0.6 & 0.6 & 0.5 & 0.5 & 0.5 & 0.4 & 0.35 & 0.008 & 0 & 0 & 0.65 & 0.3 & 0.5 & 0.4 & 1.3 \\
	\multicolumn{1}{l|}{19} & 2.1 & 1.8 & 1.6 & 1.6 & 1.2 & 1.1 & 1 & 1 & 1 & 1 & 0.8 & 0.75 & 0.7 & 0.7 & 0.7 & 0.6 & 0.55 & 0.23 & 0 & 0 & 0.7 & 0.55 & 0.7 & 0.6 & 1.3 \\
	\multicolumn{1}{l|}{20} & 1.9 & 1.7 & 1.4 & 1.4 & 1.2 & 1 & 0.85 & 0.9 & 0.9 & 0.9 & 0.65 & 0.65 & 0.6 & 0.65 & 0.5 & 0.5 & 0.45 & 0.65 & 0.65 & 0.7 & 0 & 0.85 & 0.95 & 0.85 & 1.1 \\
	\multicolumn{1}{l|}{21} & 1.8 & 1.8 & 1.6 & 1.5 & 1.2 & 1.1 & 1 & 1 & 0.95 & 1 & 0.8 & 0.8 & 0.7 & 0.75 & 0.75 & 0.65 & 0.65 & 0.35 & 0.3 & 0.55 & 0.85 & 0 & 0.19 & 0.094 & 1.5 \\
	\multicolumn{1}{l|}{22} & 2 & 2 & 1.8 & 1.7 & 1.4 & 1.3 & 1 & 1.2 & 1.1 & 1.2 & 1 & 1 & 0.85 & 0.9 & 0.95 & 0.85 & 0.85 & 0.5 & 0.5 & 0.7 & 0.95 & 0.19 & 0 & 0.13 & 1.6 \\
	\multicolumn{1}{l|}{23} & 1.9 & 1.9 & 1.7 & 1.6 & 1.3 & 1.1 & 1.1 & 1.1 & 1 & 1.1 & 0.9 & 0.9 & 0.75 & 0.8 & 0.85 & 0.75 & 0.7 & 0.4 & 0.4 & 0.6 & 0.85 & 0.094 & 0.13 & 0 & 1.5 \\
	\multicolumn{1}{l|}{24} & 2.6 & 2.4 & 2.2 & 2.1 & 1.9 & 1.9 & 1.6 & 1.7 & 1.7 & 1.7 & 1.4 & 1.4 & 1.4 & 1.4 & 1.2 & 1.3 & 1.3 & 1.3 & 1.3 & 1.3 & 1.1 & 1.5 & 1.6 & 1.5 & 0 \\
		\bottomrule
	\end{tabular}
	\caption{Pedestrian walking distance (in units of km) for pairs of sites. The Site ID number is given in Table A.\ref{tab:site data}. Site ID numbers at the left-hand side column mean the departure site, while Site ID numbers at the top row refer to the destination site.}
	\label{tab:pedestrian distance}
\end{sidewaystable}
\newpage

\section{Several Optimal Routes for Each Possible Transportation}

\begin{table}[htbp]
	\parbox{0.5\linewidth}{
		\centering
		\renewcommand{\arraystretch}{0.3}
		\begin{tabular}{ccccc}
			\hline \hline
			Route 1 & Route 2 & Route 3 & Route 4 & Route 5 \\\hline
			17 & 23 & 19 & 10 & 0 \\
			18 & 22 & 17 & 12 & 13 \\
			19 & 20 & 21 & 6 & 14 \\
			21 & 24 & 23 & 5 & 15 \\
			23 & 4 & 22 & 8 & 16 \\
			22 & 3 & 20 & 7 & 18 \\
			20 & 2 & 15 & 9 & 17 \\
			24 & 1 & 16 & 4 & 19 \\
			11 & 0 & 24 & 2 & 21 \\
			10 & 5 & 11 & 3 & 23 \\
			12 & 8 & 10 & 1 & 22 \\
			9 & 6 & 12 & 0 & 20 \\
			5 & 7 & 9 & 13 & 24 \\
			8 & 9 & 5 & 14 & 11 \\
			6 & 10 & 8 & 15 & 10 \\
			7 & 12 & 6 & 16 & 12 \\
			4 & 11 & 7 & 19 & 8 \\
			3 & 13 & 4 & 18 & 6 \\
			2 & 14 & 2 & 17 & 5 \\
			1 & 15 & 3 & 21 & 7 \\
			0 & 16 & 1 & 23 & 9 \\
			13 & 17 & 0 & 22 & 4 \\
			14 & 19 & 13 & 20 & 3 \\
			15 & 18 & 14 & 24 & 2 \\
			16 & 21 & 18 & 11 & 1 \\
			17 & 23 & 19 & 10 & 0\\
			\hline\hline
		\end{tabular}
		\caption{Optimal time (78 min) by car.}
		\label{tab:ordersCarTime}
	}
	\hfill
	\parbox{0.5\linewidth}{
		\centering
		\renewcommand{\arraystretch}{0.3}
		\begin{tabular}{ccccc}
	\hline \hline
	Route 1 & Route 2 & Route 3 & Route 4 & Route 5 \\\hline
	23 & 1 & 19 & 17 & 8 \\
	21 & 2 & 17 & 18 & 11 \\
	17 & 4 & 21 & 19 & 12 \\
	19 & 9 & 22 & 21 & 16 \\
	18 & 7 & 23 & 22 & 17 \\
	16 & 6 & 24 & 23 & 18 \\
	12 & 10 & 20 & 24 & 19 \\
	11 & 13 & 16 & 20 & 21 \\
	6 & 14 & 12 & 15 & 22 \\
	7 & 15 & 11 & 14 & 23 \\
	9 & 20 & 6 & 13 & 24 \\
	4 & 24 & 7 & 10 & 20 \\
	3 & 22 & 9 & 8 & 15 \\
	2 & 23 & 4 & 5 & 14 \\
	1 & 21 & 2 & 4 & 13 \\
	0 & 19 & 1 & 2 & 10 \\
	5 & 18 & 0 & 3 & 6 \\
	8 & 17 & 3 & 1 & 7 \\
	10 & 16 & 5 & 0 & 9 \\
	13 & 12 & 8 & 9 & 4 \\
	14 & 11 & 10 & 6 & 2 \\
	15 & 8 & 13 & 7 & 1 \\
	20 & 5 & 14 & 11 & 0 \\
	24 & 3 & 15 & 12 & 3 \\
	22 & 0 & 18 & 16 & 5 \\
	23 & 1 & 19 & 17 & 8 \\
	\hline\hline
\end{tabular}
\caption{Optimal time (115 min) on foot.}
\label{tab:ordersPedestrianTime}
	}
\end{table}

\begin{table}[htbp]	
	\renewcommand{\arraystretch}{0.3}
	  \begin{subtable}[t]{0.3\textwidth}
		\begin{center}
	\begin{tabular}{c}
	\hline \hline
	Route \\\hline
	19 \\
	18 \\
	17 \\
	23 \\
	22 \\
	21 \\
	0 \\
	1 \\
	2 \\
	3 \\
	4 \\
	5 \\
	7 \\
	9 \\
	6 \\
	8 \\
	10 \\
	12 \\
	11 \\
	13 \\
	14 \\
	15 \\
	16 \\
	20 \\
	24 \\
	19 \\
	\hline\hline
\end{tabular}

\caption{}
\label{tab:ordersBusTime}
	\end{center}
	\end{subtable}
	\quad
	  \begin{subtable}[t]{0.3\textwidth}
	\begin{center}	
		\renewcommand{\arraystretch}{0.3}
		\begin{tabular}{c}
			\hline \hline
			Route \\ \hline
			20 \\
			16 \\
			15 \\
			14 \\
			13 \\
			11 \\
			10 \\
			12 \\
			7 \\
			9 \\
			6 \\
			8 \\
			5 \\
			4 \\
			3 \\
			2 \\
			1 \\
			0 \\
			21 \\
			22 \\
			23 \\
			17 \\
			18 \\
			19 \\
			24 \\
			20 \\
			\hline\hline
		\end{tabular}
		\caption{}
		\label{tab:ordersPedestrianDistance}
	\end{center}
\end{subtable}
	\quad			
	  \begin{subtable}[t]{0.3\textwidth}
	\begin{center}		
		\renewcommand{\arraystretch}{0.3}
		\begin{tabular}{c}
			\hline \hline
			Route \\ \hline
			17 \\
			18 \\
			19 \\
			21 \\
			23 \\
			20 \\
			16 \\
			15 \\
			13 \\
			14 \\
			24 \\
			5 \\
			8 \\
			6 \\
			7 \\
			9 \\
			4 \\
			3 \\
			2 \\
			1 \\
			0 \\
			10 \\
			11 \\
			12 \\
			22 \\
			17 \\
			\hline\hline
		\end{tabular}
		\caption{}
		\label{tab:ordersCarDistance}
	\end{center}
\end{subtable}
	\caption{Optimal routes for: A.$\ref{tab:ordersBusTime}$ shortest time by bus (117 min), A.$\ref{tab:ordersPedestrianDistance}$ on foot (7.844 km), and A.$\ref{tab:ordersCarDistance}$ shortest distance by car (13.916 km).}
	\label{tab:ordersThreeTogether}
\end{table}
	\end{appendices}
\end{document}